\documentclass[twocolumn,aps]{revtex4}
\usepackage{graphicx}

\begin{document}
\title{Helstrom Theorem by No-Signaling Condition}

\author{Won-Young Hwang$^*$,
}

\affiliation{Department of Physics Education, Chonnam National
University, Kwangjoo 500-757, Republic of Korea}

\begin{abstract}
We prove a special case of Helstrom theorem by using no-signaling
condition in the special theory of relativity that faster-than-light
communication is impossible.

\noindent{PACS: 03.67.-a, 03.65.Wj}

\end{abstract}
\maketitle
Quantum bits (qubits) are fundamentally different from classical
bits in that unknown qubits cannot be copied with unit efficiency
\cite{Die82,Woo82,Yue86} (no-cloning theorem). Another related
property of qubits is that nonorthogonal qubits cannot be
distinguished with certainty \cite{Che00}.

Interestingly, however, it has been found that the no-signaling
condition is entangled with other impossibility proofs
\cite{Gis89,Gis98,Bar02,Pat03}. In particular, it has been shown
that no-signaling condition gives the same tight bound on
probability of conclusive measurement as obtained by quantum
mechanical formula \cite{Bar02}.

In this paper, we add one in the list of theorems that can be proven
by the no-signaling condition. We prove a special case of Helstrom
theorem \cite{Hel76}. Result in this paper is closely related to
other's works but different. In particular, our argument is quite
similar to the one in Ref. \cite{Gis89}. Our contribution is an
observation that violation of Helstrom theorem implies that two
appropriately chosen (different) decompositions of the same density
operator can be discriminated. This paper is organized as follows.
We describe the proposition that we will prove. We prove it by
no-signaling condition and then we conclude.

Roughly speaking, Helstrom theorem means that the more
non-orthogonal two qubits are, the more difficult it is to
discriminate them by positive-operator-valued-measurement
\cite{Che00}. Let us consider a special case of Helstrom theorem.

{\bf Proposition-1}: Consider two non-orthogonal qubits,
$|\alpha\rangle$ and $|\beta\rangle$, whose overlap, $|\langle
\alpha | \beta\rangle|^2$, is between $0$ and $1$. We are given a
qubit that is either $|\alpha\rangle$ or $|\beta\rangle$ with equal
a priori probability, $1/2$. We want to identify the qubit quantum
mechanically. Identifier of the qubit gives either an output, $0$,
or the other output, $1$. $P_E$ is the probability of making error
in the identification. Minimal value of $P_E$ is given by
$P_E^{m}=(1/2)[1-\sqrt{1-|\langle \alpha | \beta\rangle|^2}]$
\cite{Che00}. $\diamondsuit$

(Proposition-1 has interesting applications in quantum cryptography.
For example, Bennett 1992 quantum key distribution protocol
\cite{Ben92} and quantum remote gambling protocol \cite{Hwa01}.)
Before we prove Proposition-1, let us introduce the followings. Any
pure qubit $|i\rangle \langle i|$ can be represented by a
three-dimensional Euclidean Bloch vector $\hat{r}_i$ as $|i\rangle
\langle i|= (1/2)({\bf 1}+ \hat{r}_i \cdot {\bf \vec{\sigma}})$
\cite{Nie00}. Here ${\bf 1}$ is identity operator,  ${\bf
\vec{\sigma}}= (\sigma_x, \sigma_y, \sigma_z)$, and $\sigma_x,
\sigma_y, \sigma_z$ are Pauli operators. Two Bloch vectors
corresponding to $|\alpha\rangle$ and $|\beta\rangle$ are
$\hat{r}_{\alpha}$ and $\hat{r}_{\beta}$, respectively. We define an
angle between $\hat{r}_{\alpha}$ and $\hat{r}_{\beta}$ to be
$2\theta$. That is, $|\langle \alpha | \beta\rangle|^2 = \cos^2
\theta$. A pure state $|\gamma \rangle$ is defined as that its Bloch
vector $\hat{r}_{\gamma}$ bisects the two Bloch vectors
$\hat{r}_{\alpha}$ and $\hat{r}_{\beta}$ in the same plane, namely
$\hat{r}_{\gamma}= C(\hat{r}_{\alpha}+\hat{r}_{\beta})$ where $C$ is
a constant for normalization. A pure state $|\delta\rangle$ is
defined as its Bloch vector $\hat{r}_{\delta}$ makes an angle
$\pi/2$ and $\pi/2 + \theta$ with the Bloch vector
$\hat{r}_{\gamma}$ and $\hat{r}_{\alpha}$, in the same plane,
respectively. A pure state $|-\delta\rangle$ is defined as its Bloch
vector $\hat{r}_{-\delta}$ is the negative of that of
$|\delta\rangle$, namely $-\hat{r}_{\delta}$. Note that all Bloch
vectors here are in the same plane.

Let us start the proof. Consider an entangled state for Alice and
Bob who are supposed to be remotely separated usually,
\begin{equation}
\label{A} |\psi\rangle = \sqrt{p}\hspace{2mm} |0\rangle_A
|\alpha\rangle_B + \sqrt{1-p} \hspace{2mm} |1\rangle_A |\delta
\rangle_B.
\end{equation}
Here, $|0\rangle$ and $|1\rangle$ are two orthogonal qubits, $A$ and
$B$ denote Alice and Bob, and $p=\frac{1}{1+\sin\theta}$ and
$1-p=\frac{\sin\theta}{1+\sin\theta}$. If Alice performs a
measurement in $\{|0\rangle, |1\rangle \}$ basis, therefore, Bob is
given a mixture of $|\alpha\rangle \langle \alpha|$ and $|\gamma
\rangle \langle \gamma|$ with respective probabilities $p$ and
$1-p$. Then Bob's density operator $\rho_B$ is given by $\rho_B=
p\hspace{1mm}|\alpha\rangle \langle \alpha|+
(1-p)\hspace{1mm}|\gamma \rangle \langle \gamma|= (1/2) \{ {\bf 1}+
\hat{r}_B \cdot {\bf \vec{\sigma}}\}$, where $ \hat{r}_B = p
\hspace{1mm} \hat{r}_{\alpha}+(1-p)\hspace{1mm} \hat{r}_{\gamma}$.
Note that Bloch vector of a mixture is given by sum of Bloch vectors
of pure states constituting the mixture with corresponding
probabilities as weighting factors. However, theorem of
Gisin-Hughston-Jozsa-Wootters says that, with the state in Eq.
(\ref{A}), Alice can generate any decomposition of the Bob's mixture
\cite{Gis89,Hug93,Nie00} by appropriate choice of her measurement
basis. (Usually this theorem is known as that of the latter three
authors. However, the theorem had been already demonstrated by Gisin
\cite{Gis89}.) However, we have a relation that $  \hat{r}_B = p
\hspace{1mm} \hat{r}_{\alpha}+(1-p)\hspace{1mm} \hat{r}_{\gamma}= p
\hspace{1mm} \hat{r}_{\beta}+(1-p)\hspace{1mm} \hat{r}_{-\gamma}$,
which means that the density operator $\rho_B$ can also be
decomposed as $\rho_B= p \hspace{1mm}|\beta \rangle \langle \beta| +
(1-p) \hspace{1mm}|-\gamma \rangle \langle - \gamma|$. Thus the
state in Eq. (\ref{A}) can also be written as
\begin{equation}
\label{B} |\psi\rangle = \sqrt{p}\hspace{2mm} |0 ^{\prime}\rangle_A
|\beta \rangle_B + \sqrt{1-p} \hspace{2mm} |1 ^{\prime}\rangle_A
|-\delta \rangle_B,
\end{equation}
where $\{|0 ^{\prime}\rangle $, $|1 ^{\prime}\rangle \}$ is another
orthogonal basis.

Now let us assume that there exists a binary detector of any kind
whose probability of error $P_E$ is less than $P_E^{m}$ for the two
non-orthogonal states $|\alpha \rangle $ and $|\beta \rangle$. That
is, the detector gives outcomes $0$ and $1$ for $|\alpha \rangle $
and $|\beta \rangle$, respectively, with a probability $1-P_E$. Then
Alice and Bob can do faster-than-light communication in the
following way. First Alice and Bob prepare many copies of the state
in Eq. (\ref{A}). If Alice wants to send a bit $0$ (bit $1$) then
Alice performs measurements on her qubits in $\{|0 \rangle $, $|1
\rangle \}$ ($\{|0 ^{\prime}\rangle $, $|1 ^{\prime}\rangle \}$)
basis. Bob can discriminate the two cases by performing measurements
on his qubits using the detector: In the case of bit $0$ (bit $1$),
$|\alpha\rangle$ ($|\beta\rangle$) is generated with probability $p$
at Bob's site. Then $p \hspace{1mm}(1-P_E)> 1/2$ because $P_E
<P^m_E$ and $p \hspace{1mm}(1-P^m_E)=1/2$. That is, in the case of
bit $0$ (bit $1$), the detector gives outcome $0$ (outcome $1$) with
a probability larger than $1/2$. Therefore, whatever outcomes are
given for the other state, Bob can discriminate the two cases.
$\diamondsuit$


We proved a special case of Helstrom theorem, Proposition-1, by
using no-signaling condition in special theory of relativity that
faster-than-light communication is impossible.

I thank Marco Piani very much for a helpful correction.


\end{document}